\documentclass[twocolumn,amsmath,amssymb,pra]{revtex4}

\usepackage{graphicx}
\usepackage{dcolumn}
\usepackage{bm}
\usepackage{subfigure}

\begin{document}


\title {
Landau diamagnetism of the free electron gas as a Fermi surface effect
}

\author {A. V. Nikolaev}

\affiliation{
Skobeltsyn Institute of Nuclear Physics, Moscow
State University, Vorob'evy Gory 1/2, 119234, Moscow, Russia
}

\affiliation{Department of Problems of Physics and Energetics, Moscow Institute of Physics and Technology, 141700 Dolgoprudny, Russia}

\date{\today}

\begin{abstract}
The diamagnetic response of the free electron gas called the Landau diamagnetism is a complex and elusive effect requiring laborious computations.
Here based on the semi-classical treatment of the problem
I present a clear picture of the Landau diamagnetism at zero temperature, which offers a simple derivation of this effect and leads to
important consequences:
1) the diamagnetic response is due to electron states in a very narrow Fermi surface region in the $k$-space,
2) small Fermi energy oscillations in an applied magnetic field are caused by redistribution (inflow or outflow) of electrons
from the equatorial region of the Fermi surface.
The consideration is based on a structure called magnetic tube whose electron states surround a certain Landau level in $k$-space.
A completely filled magnetic tube does not change its energy in an applied magnetic field as if it complied with the Bohr -- van Leeuwen theorem.
The intersection of tubes with the Fermi surface leads to the appearance of partially occupied tubes in the region
of intersection. The reconstruction of electron states in a magnetic field in this very small narrow region
gives rise to the Landau diamagnetic response.
In addition to the Landau diamagnetism this approach fully describes the oscillatory de Haas - van Alphen contribution
to the magnetic susceptibility from the equatorial region of the Fermi sphere.
\end{abstract}



\maketitle

\section{Introduction}
\label{sec:int}

The diamagnetic effect of the free electron gas was first obtained by Landau \cite{Lan0,Lan,Lan1} in 1930,
a good historical context of the event can be found in the review of Pokrovskii \cite{Pok} devoted to the Landau heritage.
In the course of Landau and Lifshitz \cite{Lan1} the diamagnetic response in a magnetic field is computed at finite temperature $T \neq 0$ by summing up
contributions to the grand thermodynamical potential $\Omega(T,\, V,\, \mu)$ (here $\mu$ is the chemical potential, $V$ is the gas volume)
over wave vector quantum numbers $k_z$, $k_x$, and an integer $n$ numbering electronic states called Landau levels.
The sum over $n$ is then approximated by the Euler-Maclaurin summation formula \cite{Lan1}.
The calculation is rather complex from the mathematical viewpoint, and is not immediately related to the Fermi surface.
In addition, since the grand potential $\Omega$ is a function of $\mu$, there is no question about the change of the Fermi energy
in the case when the number of electrons, $N$, is fixed.

An alternative picture of the Landau diamagnetism is given in Refs.\ \cite{Pip,Pip2,Kit,Zim} 
There, at $T=0$ one considers a very thin slice of the Fermi sphere (of thickness $d k_z$)
cut normal to the direction of an applied magnetic field $\vec{H}$ (parallel to the $z-$axis).
The complexity of this approach is that depending on the kinetic energy of the translational electron movement (parallel to $\vec{H}$),
the same Landau level can be above or below the Fermi energy $E_F$, which leads to an abrupt change of the energy of the $d k_z$ slice.
Conclusions obtained from the slice can not be immediately extended to the whole picture, because
one has to resort to a procedure of averaging the effect over the slices \cite{Sho}.
In obtaining the oscillatory (de Haas - van Alphen) effect it is however pointed out
that the oscillatory contributions from all $k_z-$slices cancel each other everywhere
except in a region where the cross section normal to $\vec{H}$ passes through an extremum \cite{Pip,Pip2,Sho}.

An important contribution to the understanding of the problem of the diamagnetism of conduction electrons was made in the pioneer work of Peierls \cite{Pei}.
Within the tight-binding approximation he obtained an expression for the magnetic susceptibility consisting of three terms \cite{Pei,Wil1},
\begin{eqnarray}
     \chi = \chi_1 + \chi_2 + \chi_3 ,
\label{i1}
\end{eqnarray}
where $\chi_1$ is the diamagnetic susceptibility analogous to that of isolated metal atoms,
$\chi_2$ is a term which has no simple physical interpretation and the term $\chi_3$, which at zero temperature is given by
\begin{eqnarray}
 \chi_3 = -\frac{e^2}{48 \pi^3 \hbar^2 c^2} \int
 \left\{ \frac{\partial^2 E}{\partial k_x^2} \frac{\partial^2 E}{\partial k_y^2}
   - \left( \frac{\partial^2 E}{\partial k_x \partial k_y}  \right)^2 \right\} \frac{dS}{\nabla_k E} .  \nonumber \\
 \label{i2}
\end{eqnarray}
Here $E(\vec{k})$ is the electron band energy and the integration is taken over the Fermi surface.
The term $\chi_3$ was considered as leading in the diamagnetism of conduction electrons \cite{Pei,Wil1,Wil2}.
In principle, Eq.\ (\ref{i2}) suggests that the diamagnetic effect is due to the Fermi surface electron states.
However, the other contributions (like $\chi_1$ and $\chi_2$ in Eq.\ (\ref{i1})) spoil this picture.
More reservations in respect to Eq.\ (\ref{i2}) have been put forward in subsequent works \cite{Wil1,Wil2,Ada,KK}.
Trying to derive Eq.\ (\ref{i2}) accurately, Wilson \cite{Wil2} has found that some terms can not be explicitly evaluated and
some are not expressible in terms of derivatives of the band energy.
In a recent work of Briet {\it et al.} \cite{Bri} all these additional contributions have been obtained
at the mathematical level of accuracy. 
Briet {\it et al.} following the initial idea of Kjeldaas and Kohn \cite{KK} 
have concluded that Eq.\ (\ref{i2}) is valid only in the limit of small electron density.

Because of this complexity, the Landau diamagnetism is not considered as a Fermi surface effect.
This however can be clearly seen in the semi-classical approach presented in this paper.
It is based on a partition of the Fermi sphere in two regions.
The first big region consists of a set of many tubes of finite width and length covering almost all Fermi sphere,
but the energy of all electron states in a tube remains unchanged in an applied magnetic field.
Thus, the elementary unit here is a tube of finite width and length, Fig.\ \ref{fig0},
sandwitching a certain Landau level inside it. All electron states in the tube are occupied and diamagnetically inert.
We shall consider this structure in detail in the next section.
The second part comprises a very thin region near the Fermi surface.
The second part also consists of tubes but now they are only partially occupied and because of it, their energy in the magnetic field is changed.
The ``tube'' representation of the Fermi sphere induced by Landau levels,
facilitates the following analysis of diamagnetism and the oscillating energy contribution.

We should also mention the important question of the Fermi energy ($E_F$) change in the applied magnetic field.
This is a very weak effect, considered first in detail by Kaganov {\it et al}. \cite{Kag}.
Some consequences of the phenomenon including weak oscillations of the density of states at the Fermi level
are further discussed by Shoenberg in Ref.\ \cite{Sho}.
Below we will unveil a mechanism of this effect, which is caused by peculiarities of electron population
in the equatorial region of the Fermi sphere. Depending on the applied magnetic field there can be a small inflow or outflow of electrons
from the equatorial region to other states of the Fermi surface.

This irregular behavior is
closely related with the oscillation effect demonstrated by several physical quantities \cite{Sho}.
For the first time it was observed in the oscillations of magnetoresistance of bismuth films \cite{Shub} (Shubnikov - de Haas effect),
later -- in oscillations of the magnetic susceptibility \cite{Haa} (de Haas - van Alphen effect).
Onsager \cite{Ons} and Lifshitz \cite{Lif2,Lif} based on the semi-classical description of the
movement of an electron in a magnetic field, showed that the change in $1/H$ is determined by extremal cross-sections of the Fermi surface
in a plane normal to the magnetic field. This observation has appeared to be crucial for a deep understanding of the nature of oscillations
and allowed for a generalization in the case of arbitrary dispersion law of electrons.
A good historical and theoretical review of the effect is given in the book of Shoenberg \cite{Sho}.

Interestingly, while the de Haas-van Alphen oscillations have been closely related to the shape of the Fermi surface,
the Landau diamagnetism in general has not been perceived as a Fermi surface phenomenon.
Here we will show that the Landau diamagnetism of the free electron gas is also directly connected with the Fermi surface electron states.

The paper is written as follows: in the section \ref{sec:an} the concept of the magnetic tube is introduced, which is used in section \ref{sec:pert}
for selecting diamagnetically active tubes and electron states and
later in section \ref{sec:Lan} for computation of energy and magnetic susceptibility.
The de Haas-van Alphen effect is considered in section \ref{sec:dHvA}.
Here the analytical calculations become more difficult because one has to deal with several different cases.
Nevertheless, in all these cases the physical picture remains clear and transparent.
Our conclusions are summarized in section \ref{sec:con}.
Simple integrals over infinitesimal $(k_x,k_z)$-cross-sections in the $\vec{k}$-space used in this paper are explicitly given
in Appendices of the Supplementary Materials.

\section{Tubes in $\vec{k}$-space and their properties}
\label{sec:an}

In an external magnetic field $\vec{H}$ directing along the $z$-axis,
the energy of the electron is given by \cite{Lan0,Lan,Lan1}
\begin{eqnarray}
     E = \hbar \omega \left( n+\frac{1}{2} \right) + \frac{\hbar^2 k_z^2}{2 m} ,
\label{l0}
\end{eqnarray}
where $n$ is integer (numbering the Landau levels), $k_z$ is the $z-$component of the wave vector $\vec{k}$,
and the cyclotron frequency
\begin{eqnarray}
     \omega = \frac{eH}{mc} .
\label{l4}
\end{eqnarray}
Here $m$ and $e$ are the electron mass and charge; $c$ is the speed of light.
In correspondence with Eq.\ (\ref{l0}) the energy of electron is presented by two contributions,
the contribution $E_{\perp}$ from the movement in the plane, perpendicular to $\vec{H}$ (i.e. in the plane $(k_x,k_y)$)
and the contribution $E_z = \hbar ^2 k_z^2 /2m$ from the movement parallel to $\vec{H}$ (i.e. along the $z$-axis).
In the following we consider only the component  $E_{\perp}$, because the parallel component $E_{\parallel}=E_z$
is unchanged in the magnetic field.

Below we will follow the widely used semiclassical representation of electron orbits in momentum space \cite{Ons,Lif,Pip,Zim,Sho},
when the movement in the magnetic field in the $(k_x,k_y)$ plane is described by a quantized orbit
(although the variable $k_x$, $k_y$ are no longer `good quantum numbers'), $\oint \vec{p}\, d \vec{r} = 2\pi \hbar\, ( n + \gamma)$,
which corresponds to the $n$-th Landau level. For the parabolic energy law which is the case for free electrons,
$\gamma = 1/2$ \cite{Rot,Sho}, which accounts also for the zero point energy.
At each value of $k_z$, the quantized orbits are circles in the $(k_x, k_y)$ plane whose area is
\begin{eqnarray}
 S_n =  \frac{2\pi e H}{ c \hbar } \left(n + \frac{1}{2} \right)
\label{l1}
\end{eqnarray}
and the energy is given by
\begin{eqnarray}
     E_n = \hbar \omega \left( n+\frac{1}{2} \right) .
\label{l3}
\end{eqnarray}

\begin{figure}[ht]\center
\begin{tabular}{l r}
\includegraphics[width=30mm]{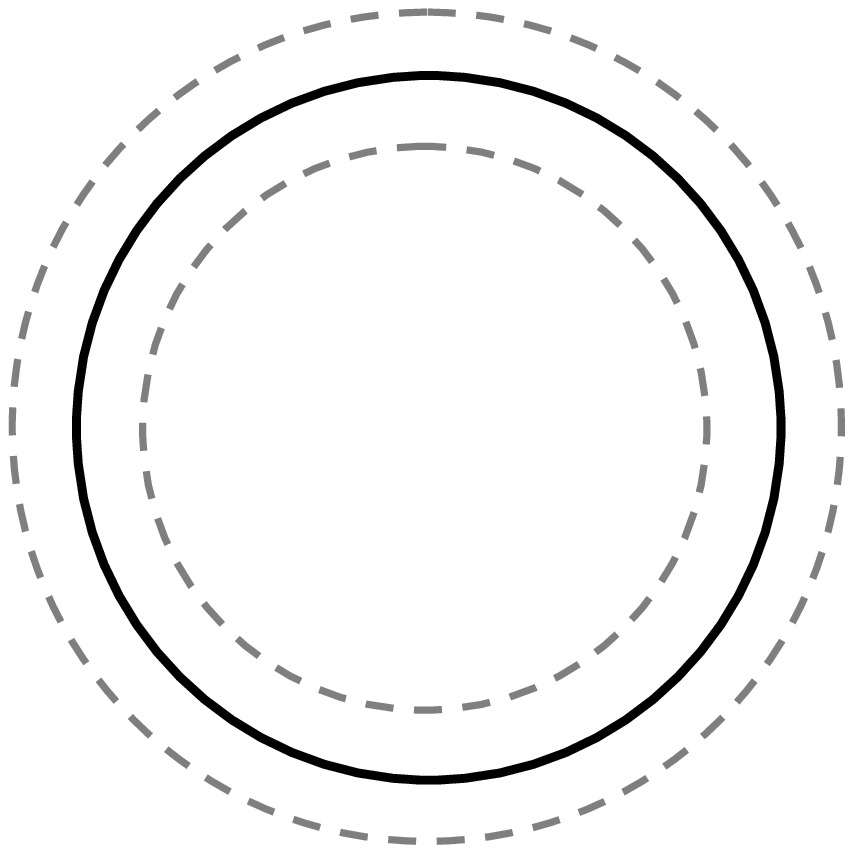}
&
\includegraphics[width=30mm]{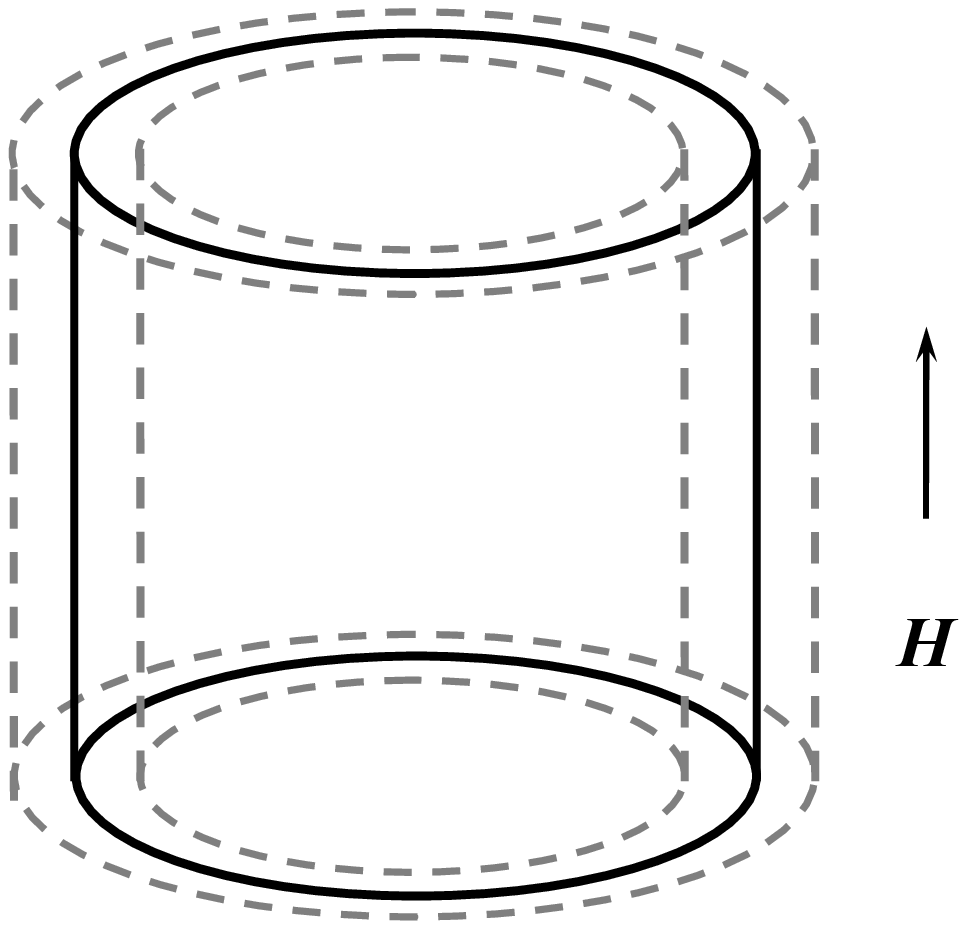}
\end{tabular}
%
\caption{Magnetic tube and the corresponding Landau level.
On the left: the $(k_x,k_y)$ tube cross-section and the $n$-th Landau orbit (the solid circle with the in-plane energy $E_n$).
The dashed circles correspond to the auxiliary orbits with energies $E_n^{aux}$ and $E_{n+1}^{aux}$.
On the right: the tube in the $\vec{k}$-space.
Without magnetic field electron states are distributed throughout the whole tube, in the presence --
only on the Landau orbit in the middle.
}
 \label{fig0}
\end{figure}

It is well known that the average density of electron states in the $\vec{k}$-space remains the same
as without magnetic field.
To understand better the reconstruction of the electron structure in the magnetic field $H$,
we select in the $\vec{k}$-space a tube, whose number of electron states and the energy of all states
do not change in the presence of $H$.
For that we consider auxiliary electron orbits of the area
\begin{eqnarray}
       S_n^{aux} = \frac{2\pi e H}{ c \hbar }\, n
\label{l5a}
\end{eqnarray}
with energies
\begin{eqnarray}
     E_n^{aux} = \hbar \omega n .
\label{l5}
\end{eqnarray}
Note the the $n$-th Landau orbit defined by Eqs.\ (\ref{l1}), (\ref{l3}) is situated between the auxiliary orbits $n$ and $n+1$,
Fig.~\ref{fig0} left panel,
and its energy $E_n$ lies between $E_n^{aux}$ and $E_{n+1}^{aux}$.
Below we show that the number of electron states with energies $E_n^{aux} \leq E \leq E_{n+1}^{aux}$ without field
equals the number of electron states condensing on the $n$-th Landau level in the presence of the field.
The same holds for their total energies.

For that we calculate the density of electron states ${\cal N}_{\perp}$ in the $(k_x,k_y)$-plane,
\begin{eqnarray}
     {\cal N}_{\perp} = \frac{d N_{\perp} }{d E_{\perp} } = \frac{2m}{\hbar^2} \frac{L_x L_y}{2\pi} .
\label{l9}
\end{eqnarray}
and notice that ${\cal N}_{\perp}$ is independent of the energy $E_{\perp}$.
(Here $L_x$, $L_y$ and $L_z$ are distances of the free electron gas box in $x$, $y$ and $z$ directions, respectively.)
Using (\ref{l9}),
we find the number of electron states in the $n$-th tube without field, i.e.
in the energy range $E_n^{aux} \leq E \leq E_{n+1}^{aux}$,
\begin{subequations}
\begin{eqnarray}
     \triangle N_n(H=0) = \int_{E_n^{aux}}^{E_{n+1}^{aux}} {\cal N}_{\perp}\, dE_{\perp} = {\cal N}_{\perp}\, \hbar \omega = 2 N_p . \quad
\label{l10a}
\end{eqnarray}
Here $N_p$ is the number of space electron states (without spin polarization) on the $n$-th Landau level in the applied magnetic field ($H \neq 0$),
\begin{eqnarray}
     N_p = \frac{L_y}{2\pi}\, \frac{m\omega}{\hbar} L_x .
\label{l8}
\end{eqnarray}
Calculating the total energy of these states without field,
\begin{eqnarray}
   & &{\cal E}_n(H=0) = \int_{E_n^{aux}}^{E_{n+1}^{aux}} {\cal N}_{\perp} E_{\perp} dE_{\perp}   \nonumber \\
   & & =  \frac{1}{2}\left[ (E_{n+1}^{aux})^2 - (E_n^{aux})^2 \right] {\cal N}_{\perp}
   = E_n \triangle N_n , \quad
\label{l10b}
\end{eqnarray}
\end{subequations}
we find that it coincides with the energy of all electron tube states condensed on the $n$-th Landau level in the presence of the field.
Thus, we have proven that
\begin{subequations}
\begin{eqnarray}
  & &   \triangle N_n(H=0) = \triangle N_n( H \neq 0) = 2 N_p , \label{l11a} \\
  & &   {\cal E}_n(H=0) = {\cal E}_n(H \neq 0) = \hbar \omega \left( n+\frac{1}{2} \right) 2 N_p . \quad \quad \label{l11b}
\end{eqnarray}
\end{subequations}

Our consideration has been limited by the $(k_x,k_y)$ plane. However, since the movement along the $z$ axis is unchanged,
we can extent it to the three dimensional $\vec{k}$-space and define there a tube $n$, Fig.\ \ref{fig0}.
Without field, the tube contains all electron states which satisfy the inequalities $E_n^{aux} \leq E_{\perp} \leq E_{n+1}^{aux}$
and $E_z^{(1)} \leq E_z \leq E_z^{(2)}$, while in the presence of the magnetic field the states condense on
the $n$th Landau level, that is, $E_{\perp} = E_n$ and $E_z^{(1)} \leq E_z \leq E_z^{(2)}$, Fig.\ \ref{fig0}.
The upper $E_z^{(1)}$ and lower $E_z^{(2)}$ boundary of a tube can be taken arbitrary.
In practice, they are defined by intersection with the Fermi surface, Fig.\ \ref{fig1}.
We then consider two $n$-th tubes: the first tube lies completely inside the Fermi sphere and is not exposed to diamagnetism,
while the second tube containing a part of the Fermi sphere and occupied electron states below it,
is only partially filled, which results in a diamagnetic response.
We consider this effect in the following sections.
%
%
%
%

It is also worth noting that the $\vec{k}$-space partitioning depends on the value of the magnetic field, since $\omega \sim H$,
and the tube boundaries are defined by $\omega$, Eq.\ (\ref{l5}).

\section{Diamagnetically active electron states in the neighborhood of the Fermi surface}
\label{sec:pert}

\begin{figure}
\resizebox{0.4\textwidth}{!} {
\includegraphics{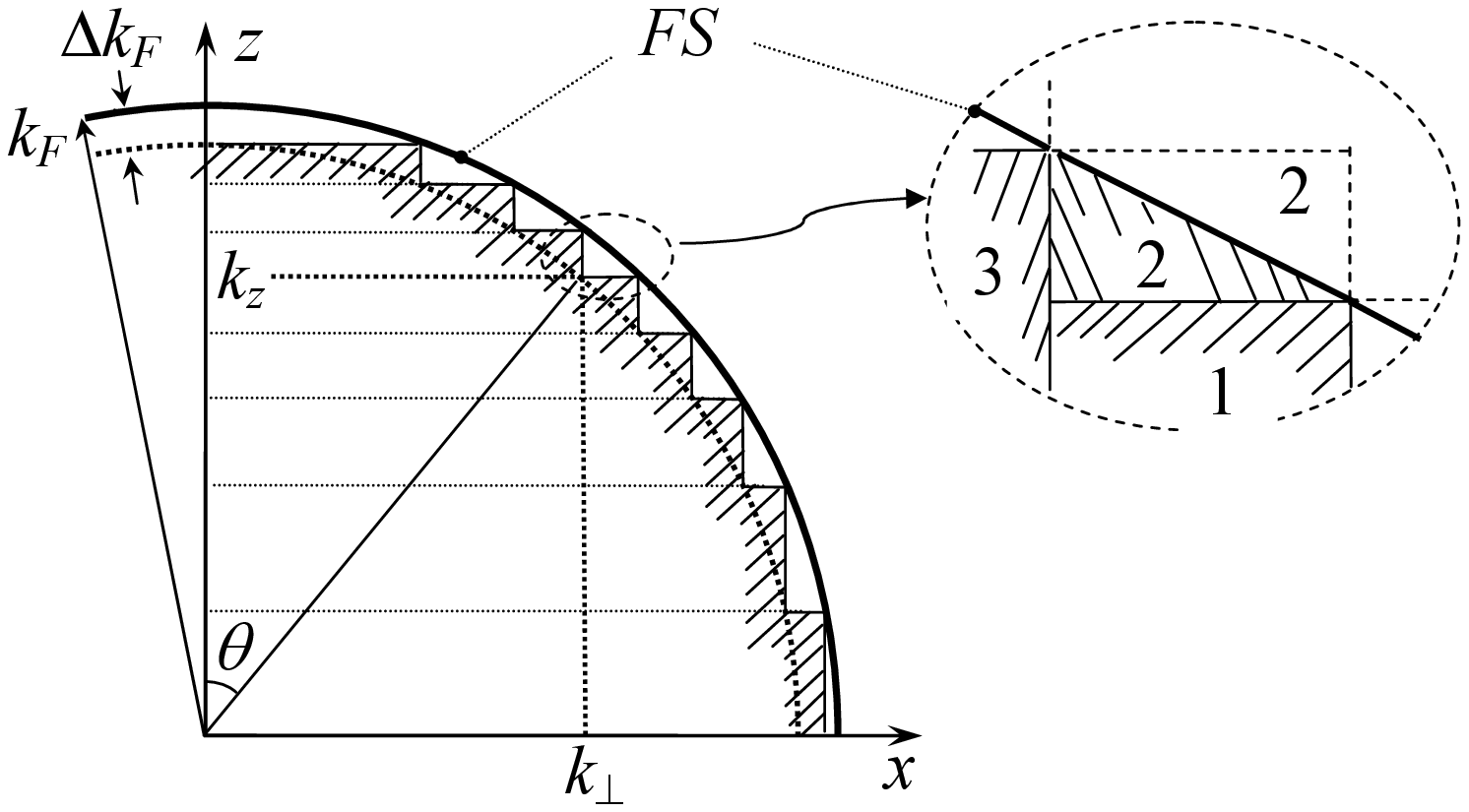}   }

\vspace{2mm}
\caption{
Magnetic tubes and the Fermi surface ($FS$), the $(k_x,k_z)$-cross-section in the $\vec{k}$-space.
The tubes inside the Fermi sphere shown as dashed area on the left panel, are completely filled
and diamagnetically inert.
Inset: 1 - completely occupied tube $n$, 2 - partially occupied tube $n$, 3 - completely occupied tube $(n-1)$.
} \label{fig1}
\end{figure}

\begin{figure}
\resizebox{0.25\textwidth}{!} {
\includegraphics{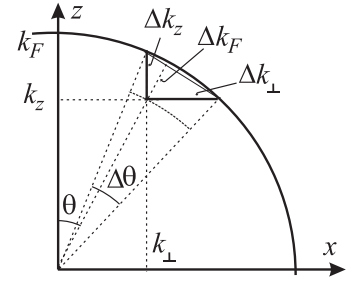}   }

\caption{
The $(k_x,k_z)$-cross-section of a partially occupied tube near the Fermi surface, see text for details.
The triangle size is greatly exaggerated, since $\triangle k_F = m \omega/\hbar k_F \ll k_F$.
} \label{fig2}
\end{figure}

Consider the Fermi surface and define necessary magnetic tubes parallel to the $z$-axis (in the direction of the magnetic field $H$), Fig.\ \ref{fig1},
as discussed in Sec.\ \ref{sec:an}.
Boundary conditions defined by in-plane circular orbits, Eq.\ (\ref{l5a}), specify a set of concentric cylindrical surfaces,
which intersect the Fermi surface in circles perpendicular to the $z$-axis.
We then draw the planes of the circles and use them to construct a set of tubes, limited by the planes and the cylindrical surfaces, which lie
inside the Fermi sphere.
The $(k_x, k_z)$ cross-section of these tubes is schematically shown in Fig.\ \ref{fig1}.
The fully occupied tubes are shown as dashed area.
The electron states of the completely filled tubes do not change their energy in a magnetic field.
Therefore, the whole effect is due to the states lying in the partially occupied tubes.
Their cross-sections in the $(k_x, k_z)$-plane look like a chain of triangles, Fig.\ \ref{fig1} and \ref{fig2}.

%
%
Consider a typical partially occupied tube, whose triangle cross-section in the $(k_x, k_z)$-plane is shown in Fig.\ \ref{fig2}.
We denote two legs of the triangle by $\triangle k_{\perp}$ and $\triangle k_z$.
Taking into account that the area of the $(k_x, k_y)$ cross-section of the $n$-th tube is $\triangle S =  S_{n+1}^{aux} - S_n^{aux} = 2\pi e H/ c \hbar$,
and that $\hbar \omega \ll E_F$, we obtain
\begin{eqnarray}
 \triangle k_{\perp} =  \frac{\triangle S}{2\pi k_{\perp}} =  \frac{m \omega}{\hbar k_F} \frac{1}{\sin \Theta} .
 \label{f3}
\end{eqnarray}
(Here $\Theta$ is the polar angle, Fig.\ \ref{fig1}.)
Therefore, the narrow surface region of the partially occupied tube is defined by
the wave vector $\triangle k_F$ shown in Figs.\ \ref{fig1} and \ref{fig2},
\begin{eqnarray}
 \triangle k_F  = \triangle k_{\perp} \sin \Theta = \frac{m \omega}{\hbar k_F}  .
 \label{f4}
\end{eqnarray}
It is remarkable that $\triangle k_F$ is independent of $\Theta$.
Therefore, the radius $k_F - \triangle k_F$ determines an auxiliary internal sphere in the $\vec{k}$-space,
which can be used for
drawing the step-wise line shown in Figs.\ \ref{fig1} and \ref{fig2},
separating the fully occupied tubes from the partially occupied ones.

Now we find the number of active electron states in the partially occupied tubes,
\begin{eqnarray}
  N = \sum_{i=1}^M \triangle N_i ,
 \label{f11}
\end{eqnarray}
where $\triangle N_i \equiv \triangle N(\Theta_i,\triangle \Theta_i)$ is the number of active states
in the $i$-th partially filled tube, and
\begin{eqnarray}
   \triangle \Theta = \frac{\triangle k_F}{k_F} \frac{1}{\cos \Theta \sin \Theta} . \label{f10b}
\end{eqnarray}
Using the infinitesimal property of the $(k_x, k_y)$ cross-section (details are given in Supplementary Materials, Appendix~A)
we find
\begin{eqnarray}
   \triangle N = 2\rho \triangle V = 2\pi \rho k_F (\triangle k_F)^2 \frac{1}{\cos \Theta} , \label{f10a}
\end{eqnarray}
where $\rho = V/(2\pi)^3$.
Notice, that Eq.\ (\ref{f10a}) has a singularity at $\Theta = \pi/2$, and Eq.\ (\ref{f10b}) at $\Theta = 0$ and $\pi/2$.
Therefore, the polar and equatorial region of the Fermi sphere should be considered more attentively,
see the Supplementary Materials.
The equatorial region is thoroughly discussed in Sec.\ \ref{sec:dHvA} below.

Since for usual magnetic fields $\triangle k_F \ll k_F$,
in Eq.\ (\ref{f11}) we can substitute the summation with the integration,
\begin{eqnarray}
  N = \int d N = \int_0^{\pi} \frac{\triangle N}{\triangle \Theta} d \Theta  .
 \label{f12a}
\end{eqnarray}
Using (\ref{f10a}), (\ref{f10b}) we arrive at
\begin{eqnarray}
  N = 4 \pi \rho k_F^2 \triangle k_F .
 \label{f15}
\end{eqnarray}
Therefore, $N \sim \triangle k_F \sim \omega \sim H$.
Since the perturbation energy for each electron state can be estimated with $\hbar^2 k_F \triangle k_F/m =\hbar \omega \sim H$,
the total energy change in the magnetic field $\sim H^2$, which leads to the constant magnetic susceptibility $\chi$.
(The rigorous computation of $\chi$ is given in the next section.)
%
%

\section{Landau diamagnetic susceptibility}
\label{sec:Lan}

%
\begin{figure}
\resizebox{0.4\textwidth}{!} {
\includegraphics{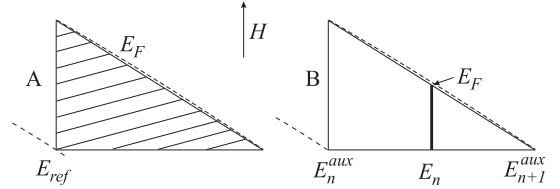}   }

\vspace{2mm}
\caption{
The $(k_x,k_z)$-cross-section of a partially occupied tube near the Fermi level.
On the left ($A$): dashed area -- occupied electron states without magnetic field, $\vec{H}=0$.
On the right ($B$): bold vertical line -- the occupied electron states (1D gas) in the magnetic field $\vec{H} \neq 0$.
$E_F$ is the Fermi energy of the one-dimensional electron gas (along the $z$-axis), whose transverse energy $E_n$
is determined by the $n$-th Landau energy.
} \label{fig4}
\end{figure}
%

A remarkable property of partially occupied tubes near the Fermi surface is that
application of a magnetic field does not lead to electron transitions between different tubes
(with the exception of a small number of electrons in the equatorial region).
Therefore, upon applying the field, there is a redistribution of electron states only within each
partially filled tube.

To demonstrate this, we consider in detail the transformation of electron states in the tube
when the magnetic field is switched on.
(Without field the number of electron states $\triangle N$ is given by Eq.\ (\ref{f10a}).)
The occupation of electron states for two cases ($H = 0$ and $H \neq 0$) is shown schematically in Fig.\ \ref{fig4}.
When $H \neq 0$, all electrons of the tube are on the $n$-th Landau level with the transverse energy $E_n$,
and occupy the lowest $k_z$-states, as shown in Fig.\ \ref{fig4}.
If all electrons remain in the tube, then the highest energy level with the wave vector
$\delta k_z$ [in respect to $k_z(\Theta)$, Fig.\ (\ref{fig2})] is found from the following relation:
\begin{eqnarray}
   2 \delta k_z N_p \rho_z = \triangle N .
 \label{e12}
\end{eqnarray}
We recall that $N_p$ is the in-plane (or transverse) folding of the $n$-th Landau level, Eq.\ (\ref{l8}),
while $\rho_z=L_z/2\pi$ is the density of electron states along $k_z$.
Substituting for $N_p = 2\pi \rho_{\perp} k_{\perp} \triangle k_{\perp} = 2\pi \rho_{\perp} k_F \triangle k_F$,
and Eq.\ (\ref{f10a}) for $\triangle N$, we obtain
\begin{eqnarray}
   \delta k_z = \frac{1}{2} \frac{\triangle k_F}{\cos \Theta} = \frac{1}{2} \triangle k_z .
 \label{e12a}
\end{eqnarray}
(The result for the infinitely small triangle cross-section can be foreseen from the geometrical reasons.)

Eq.\ (\ref{e12a}) leads to an important consequence. The energy of the highest occupied electron level coincides with
$E_F$ and the wave vector $k_F$ lies on the Fermi surface even in the applied magnetic field $H \neq 0$.
Since the conclusion holds for all partially occupied tubes (with the exception of few equatorial tubes),
the highest energy level $E_F$ is conserved as the Fermi level for all tubes and there are no electron transitions
between tubes (the only exception is the equatorial region considered later in Sec.\ \ref{sec:dHvA}.)
Therefore, in the following on applying the magnetic field we can calculate the energy change for each tube separately.

Keeping in mind that there are two energy contributions: $E_{\perp}$ in the $(k_x,k_y)$ plane and $E_z$ along the $z$ axis (parallel to the field $H$),
we obtain for the energy change of the partially filled tube in the magnetic field,
\begin{eqnarray}
  & &\triangle E = \triangle E^{H \neq 0} - \triangle E^{H = 0}  \nonumber \\
  & &= \triangle E_{\perp}^{H \neq 0} + \triangle E_z^{H \neq 0} - \triangle E_{\perp}^{H = 0} - \triangle E_z^{H = 0} . \quad
 \label{e2}
\end{eqnarray}
The notation $\triangle$ on the right hand side is an indication that the corresponding energy refers to the partially filled tube
characterized by the polar angle $\Theta$ and the angle $\triangle \Theta$, as shown in Fig.\ \ref{fig2} (see also Fig.~1 of the Supplementary Materials).
The quantities $\triangle E_{\perp}^{H=0}$, $\triangle E_z^{H=0}$,
$\triangle E_{\perp}^{H \neq 0}$ and $\triangle E_z^{H \neq 0}$, therefore refer to two energy components of the tube with and without magnetic field.

The detailed simple calculations of all components are performed in Appendix A of the Supplementary Materials
(in respect to the energy $E_{ref}$, Fig.\ \ref{fig4}).
As a result, we get
\begin{eqnarray}
  \frac{\triangle E_{\perp}^{H=0}}{\triangle N} = \frac{\triangle E_{z}^{H=0}}{\triangle N}
  = \frac{1}{3} \frac{\hbar^2}{m} k_F \triangle k_F = \frac{1}{3} \hbar \omega  \nonumber \\
 \label{ee10}
\end{eqnarray}
for the energy components without magnetic field and
\begin{subequations}
\begin{eqnarray}
  \frac{\triangle E_{\perp}^{H \neq 0}}{\triangle N} = \frac{1}{2} \hbar \omega ,  \label{ee11a} \\
  \frac{\triangle E_z^{H \neq 0}}{\triangle N} = \frac{1}{4} \hbar \omega          \label{ee11b}
\end{eqnarray}
\end{subequations}
in the applied magnetic field.
In fact, the right hand sides of Eqs.\ (\ref{ee10}), (\ref{ee11a}) and (\ref{ee11b})
represent average values of energy component independent on the tube under consideration.
The substitutions of (\ref{ee10}), (\ref{ee11a}) and (\ref{ee11b}) in (\ref{e2}) yields
\begin{eqnarray}
  \triangle E = \frac{1}{12} \hbar \omega\, \triangle N > 0 .
 \label{ee18}
\end{eqnarray}

Note that Eq.\ (\ref{ee18}) refers to any partially filled tube.
Therefore, making the summation over all tubes (as discussed in Sec.\ \ref{sec:pert})
we find
\begin{eqnarray}
  E = \frac{1}{12} \hbar \omega\, N = \frac{1}{3}\, \pi \rho \, k_F^2 \triangle k_F\, \hbar \omega .
 \label{e19}
\end{eqnarray}
Substituting (\ref{f4}) for $\triangle k_F$ in (\ref{e19}),
we calculate the magnetic susceptibility $\chi$:
\begin{eqnarray}
  \chi = -\frac{d^2E(H)}{dH^2} = -\frac{2}{3}\, \pi \rho \, k_F \frac{e^2}{mc^2} =-\frac{e^2 k_F V}{12 \pi^2 mc^2} . \nonumber \\
 \label{e20}
\end{eqnarray}
This is the celebrated expression, obtained by Landau for the diamagnetic susceptibility of the free electron gas.

\section{Equatorial contribution and oscillations of energy and magnetic susceptibility}
\label{sec:dHvA}

Earlier (Sec.\ \ref{sec:Lan}) we have obtained the diamagnetic effect based of calculations of the energy of active electrons
in the partially occupied tube of general form.
The $(k_x,k_z)$ cross-section of such a tube is shown in Figs.\ \ref{fig1}, \ref{fig2}, \ref{fig4}.
Deviations from the general situation are possible for boundary cases, which are the polar region ($\Theta=0$) with the Landau level $n=0$,
and the equatorial region ($\Theta=\pi/2$).
In Appendix B of the Supplementary Materials we analyse the polar region and show that it complies with the general case.
In the equatorial region however the situation is different.
%
\begin{figure}
\resizebox{0.2\textwidth}{!} {
\includegraphics{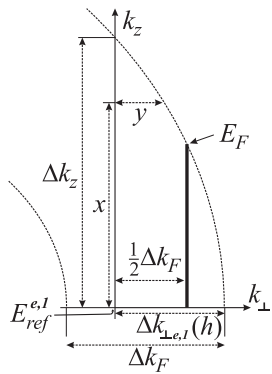}   }

\vspace{2mm}
\caption{
First equatorial tube.
The Landau level (bold line) is occupied if $h > \triangle k_F/2$,
and empty if $h < \triangle k_F/2$.
The function $y(x)$ is used for calculations.
} \label{fig5}
\end{figure}
%

The problem is that the step-wise line shown in Fig.\ \ref{fig1} can terminate at the equatorial point with $\Theta=\pi/2$
in any place with $k_{\perp,e}$ lying in the interval $k_F - \triangle k_F \le k_{\perp,e} \le k_F$,
and the equatorial point does not necessarily lie on the internal sphere of the radius $k_F - \triangle k_F$,
which is the case for all other tubes, Fig.\ \ref{fig5}.
This equatorial tube is truncated because its upper energy boundary  $E_{n+1}^{aux}$, defined by (\ref{l5}),
in general lies outside the $(k_x,k_y)$ equatorial cross-section and the Fermi sphere.
We define this irregular tube with $k_{\perp} \ge k_{\perp,e}$ as first equatorial tube.
Notice that when $k_{\perp,e}$ approaches $k_F$, the $(k_x,k_z)$ cross-section of the first equatorial tube converges to zero.
In such a situation one has to resort to the preceding tube (that is, with $k_{\perp} < k_{\perp,e}$),
which also make a small irregular contribution to the total energy.
We define it as second equatorial tube.
The other tubes follow the general dependencies considered in Sec.\ \ref{sec:Lan}.

For the first equatorial tube we define $\triangle k_{\perp,e,1} = k_F - k_{\perp,e}$, Fig.\ \ref{fig5}.
The subscript $e,1$ here and below is used to emphasise that the parameter refers to the
first equatorial tube. For the second equatorial tube we shall use the subscript $e,2$.
As we discussed above,  $\triangle k_{\perp,e,1}$ ranges from 0 to $\triangle k_F$.
In the following we shall use a short notation $h = \triangle k_{\perp,e,1}$.
Consider the important dimensionless parameter
\begin{eqnarray}
  r = \frac{h}{\triangle k_F} ,
 \label{q2}
\end{eqnarray}
determining the irregularity of the first equatorial tube. Clearly, $0 < r < 1$.
Note, that by varying $\vec{H}$ we change the structure of all magnetic tubes, and, consecutively
the parameter $r$, which is defined by the geometry of the last tube.
Therefore, $r$ implicitly depends on  $H$. In Appendix C1 of the Supplementary Materials we show
that in a first approximation $r$ is proportional to $1/H$.

For $\triangle k_{z,e,1}$ we obtain
\begin{eqnarray}
  \triangle k_{z,e,1} = \sqrt{2 k_F h} = \sqrt{2 r k_F \triangle k_F} .
 \label{q1}
\end{eqnarray}
Calculating the number of states in the first equatorial tube without magnetic field (see Appendix C1), we find
\begin{eqnarray}
  \triangle N_{e,1}^{H=0} = \frac{8 \pi r}{3} \rho \, k_F \triangle k_F \triangle k_{z,e,1} .
 \label{qq5}
\end{eqnarray}
Earlier, based on the analysis of Cornu spiral sum, Pippard estimated that the relative weight
of the extremal region should be $\triangle N_e / N \sim \sqrt{H}$ [Eq.\ (33) of \onlinecite{Pip}].
This conclusion is in agreement with Eq.\ (\ref{qq5}) since
$\triangle N_{e,1}^{H=0}/N \sim \triangle k_{z,e,1} \sim \sqrt{\omega} \sim \sqrt{H}$.

Notice that already in obtaining $\triangle N_{e,1}^{H=0}$ we have a deviation from the general case, since
\begin{eqnarray}
  \frac{\triangle N_{e,1}^{H=0}}{\triangle \Theta_{e,1}} = \frac{4r}{3} \frac{\triangle N}{\triangle \Theta} .
 \label{qq7}
\end{eqnarray}
(we recall that $\triangle \Theta_{e,1}$ defines the angular span of the first equatorial triangle in the $(k_x,k_z)$-cross-section.)
Deviations are also present for the transverse and parallel energy contribution (without field, $H=0$),
Eqs.\ (C4a) (C4b) in Appendix C1 of the Supplementary Materials.

Now we consider the situation in the magnetic field $\vec{H} \neq 0$, parallel to the $z$-axis.
We start as in Sec.\ \ref{sec:Lan} with finding the wave vector $\delta k_z$ of the highest occupied
electron state along the $z$-axis under assumption that all electrons belonging to the first equatorial tube
do not leave it.
By means of (\ref{e12}) we get
\begin{eqnarray}
  \delta k_{z,e,1} = \frac{2}{3}\, r \, \triangle k_{z,e,1} .
 \label{q9}
\end{eqnarray}
Now however the energy of the highest occupied state in general differs from $E_F$, and
therefore from the energy of highest occupied states in other tubes, Eq.\ (\ref{e12a}).
Below we consider the situation for two different cases: $0 \le r < 1/2$ (case $a$) and $1/2 \le r < 1$ (case $b$).

In the case $a$ the energy of the Landau level of the first equatorial tube $E_n - E_{ref}^{e,1} = \hbar \omega/2$, Fig.\ \ref{fig5},
is higher than $E_F$ even at $k_z = 0$.
Therefore, all electrons from this tube move to other tubes where they occupy free states above $E_F$.
As a result, a small rise in $E_F$ should occur, but since $\triangle N_{e,1} \ll N$, it is of the order of
$\hbar \omega \, \triangle N_{e,1} / N \ll \hbar \omega$.
Since $E_F - E_{ref}^{e,1} = r \hbar \omega$, the energy of the promoted electrons is
$\triangle E^{H \neq 0,a}/\triangle N_{e,1}^{H=0} = r \, \hbar \omega$
(in respect to $E_{ref}^{e,1}$).

\begin{figure}
\resizebox{0.4\textwidth}{!} {
\includegraphics{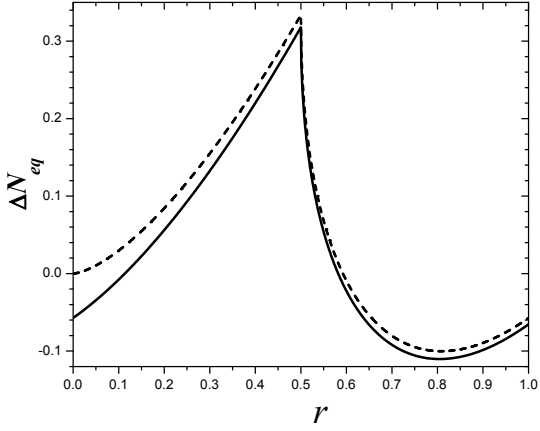}   }

\caption{
The number of electrons (in units of $8 \pi \rho (k_F \triangle k_F)^{3/2}$) promoted from the equatorial range to other tubes,
$\triangle N_{eq}=\triangle N_{e}^{H = 0} - \triangle N_{e}^{H \neq 0}$ in the applied magnetic field $H$, expressed
in terms of the dimensionless parameter $r \sim 1/H$ (see text).
Negative values imply that electrons move to the equatorial tube.
The dashed line stands for the contribution from the first equatorial tube, solid line -- from the first and second equatorial tubes.
The same plot (in units of $\hbar^2/m \cdot \sqrt{k_F} (\triangle k_F)^{3/2}$) describes a small oscillatory dependence
of the Fermi energy (chemical potential).
} \label{fig6A}
\end{figure}

In the case $b$ the Landau level at $k_z = 0$ lies below $E_F$ and in the magnetic field it becomes partially occupied by electrons with $k_z > 0$.
The maximal wave vector $\delta k_{F}^z$ of the highest occupied electron state lying on the Fermi sphere can be found
by requiring its energy to be equal to $E_F$,
\begin{eqnarray}
  \delta k_{F}^z = \sqrt{2 k_F \left(h - \frac{1}{2} \triangle k_F \right) }
  = \triangle k_{z,e,1} \sqrt{\frac{r - \frac{1}{2}}{r}} .
 \label{q11}
\end{eqnarray}
The number of the occupied electron states in the tube, $\triangle N_{e,1}^{H \neq 0,b}$,
is determined by
\begin{eqnarray}
  \frac{ \triangle N_{e,1}^{H \neq 0,b} }{\triangle N_{e,1}^{H=0}} = \frac{3}{2r^{3/2}} \sqrt{r - \frac{1}{2}} .
 \label{q12}
\end{eqnarray}
The condition $\triangle N_{e,1}^{H=0} > \triangle N_{e,1}^{H \neq 0,b}$ in terms of $r$ means $1/2 \le r < \sqrt{3} \sin \pi /9$,
while $\triangle N_{e,1}^{H=0} \le \triangle N_{e,1}^{H \neq 0,b}$ results in $\sqrt{3} \sin \pi /9 \le r < 1$.
Therefore, if $1/2 \le r < 0.529$, electrons from the first equatorial tube partially move to other (regular) tubes as it happens in the case $a$.
For $0.529 \le r < 1$ the opposite happens, that is a small number of electrons from all regular tubes move to the equatorial tube.
The change of the number of electrons in the equatorial region is shown in Fig.\ \ref{fig6A}.
%
%

To single out the irregular contribution from the equatorial region explicitly, we rewrite it
in the following form,
\begin{eqnarray}
   E = E_L + \triangle E_{eq} .
 \label{q19}
\end{eqnarray}
Here $E_L$ is the diamagnetic (regular) contribution, Eq.\ (\ref{e19}), and $\triangle E_{eq}$
stands for the irregular term from the equatorial region.
If only the first equatorial tube is accounted for, then $\triangle E_{eq} = \triangle E_{eq,1}$,
where
\begin{eqnarray}
 \triangle E_{eq,1} &=& \triangle E_{\perp,e,1}^{H \neq 0}   + \triangle E_{z,e,1}^{H \neq 0}
 - \triangle E_{\perp,e,1}^{H=0} - \triangle E_{z,e,1}^{H=0}  \nonumber \\
  & & + \triangle E_{pr,1} -\triangle E_{corr,1} .
 \label{q20}
\end{eqnarray}
Here $\triangle E_{pr,1}$ is the energy of the promoted electrons (transferred to or from regular tubes),
while $\triangle E_{corr,1}$ stands for the regular diamagnetic contribution,
\begin{eqnarray}
   \frac{ \triangle E_{corr,1} }{\triangle N_{e,1}^{H = 0}} = \frac{1}{16\, r} \, \hbar \omega .
 \label{q18}
\end{eqnarray}

\begin{figure}
\resizebox{0.4\textwidth}{!} {
\includegraphics{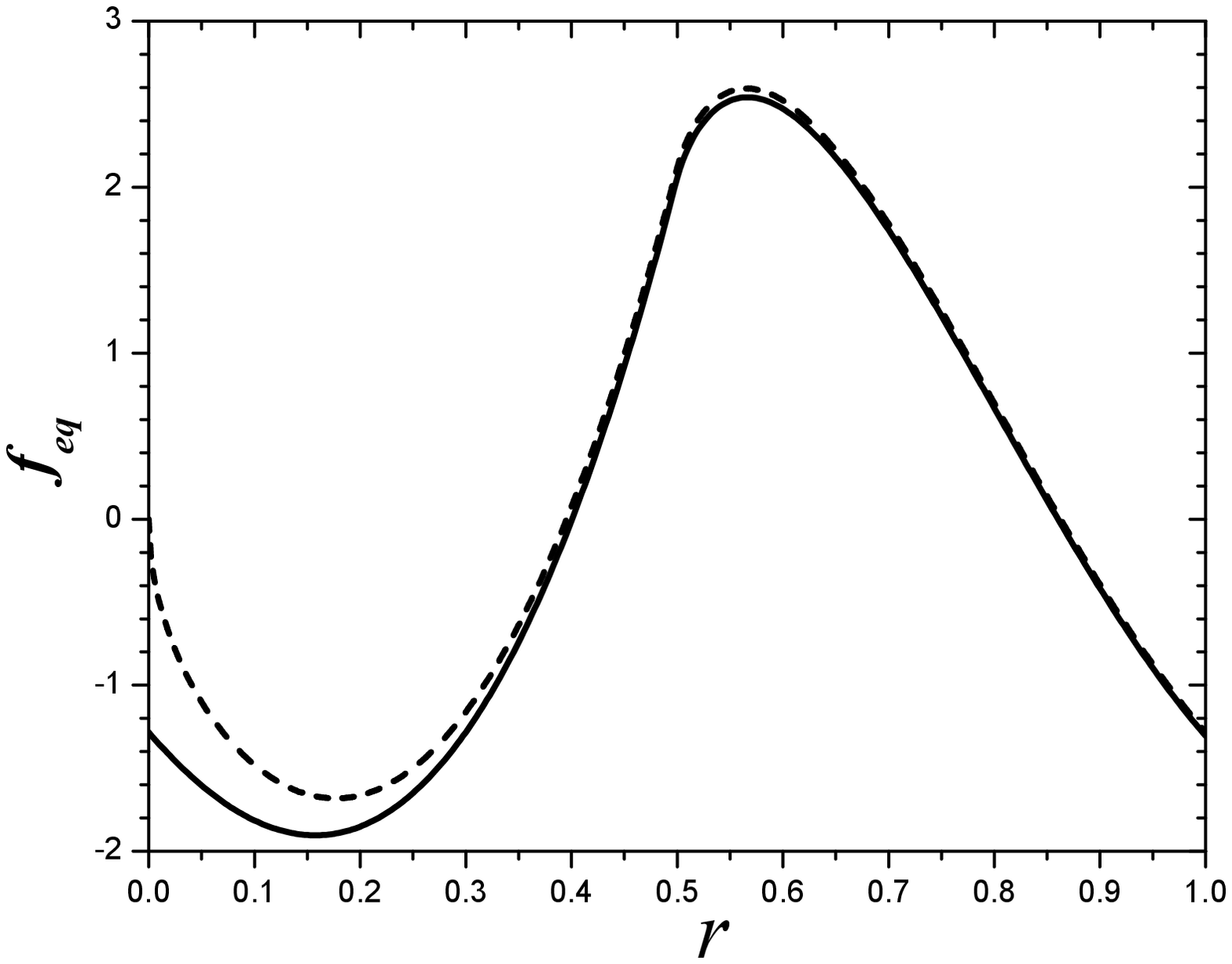}   }

\caption{
The oscillatory behavior of the irregular contribution to energy, $\triangle E_{eq} \sim f_{eq}(r)$,
from the equatorial region in the applied magnetic field $H$,
expressed in terms of the dimensionless parameter $r \sim 1/H$ (see text).
The dashed line stands for the contribution from the first equatorial tube, the solid line -- from the
first and second equatorial tubes.
} \label{fig6}
\end{figure}

Collecting all energy terms (C4a)--(C8b), written in Appendix C1 of the Supplementary Materials together,
we arrive at
\begin{eqnarray}
   \triangle E_{eq}= 2 \frac{m \pi \rho}{15 \sqrt{2}} \;
 \omega^2 \sqrt{\frac{m}{\hbar}\, \omega} \; f_{eq}(r) .
 \label{q21}
\end{eqnarray}
(The factor 2 stands for two equivalent contributions from the upper and lower Fermi semisphere.)
For the first equatorial tube we have $f_{eq}(r) = f_{eq,1}(r)$,
and the function $f_{eq,1}(r)$ has different dependences for the cases $a$ and $b$, described earlier.
In the case $a$ ($0 \le r < 1/2$)
$f_{eq,1}(r) = f_{eq,1}^a(r)$,
\begin{subequations}
\begin{eqnarray}
 f_{eq,1}^a(r) = \sqrt{r}\, (32 r^2 - 5) ,
 \label{q22a}
\end{eqnarray}
in the case $b$ ($1/2 \le r < 1$) $f_{eq,1}(r) = f_{eq,1}^b(r)$,
\begin{eqnarray}
 f_{eq,1}^b(r) = \sqrt{r}\, (32 r^2 - 5 )- 80 \left(r-\frac{1}{2}\right)^{3/2} .
 \label{q22b}
\end{eqnarray}
\end{subequations}
The dependence of $\triangle E_{eq,1} \sim f_{eq,1}(r)$ from $r$ is shown in Fig.\ \ref{fig6}.
Note that $\triangle E_{eq,1}(r=0) \neq \triangle E_{eq,1}(r=1)$,
although $r=0$ and $r=1$ refer to the same physical situation.
Below we shall see that by including two equatorial tubes, the equality of the energy at $r=0$ and $r=1$
is restored (see also Appendix C2 of the Supplementary Materials).
%
%

\begin{figure}
\resizebox{0.4\textwidth}{!} {
\includegraphics{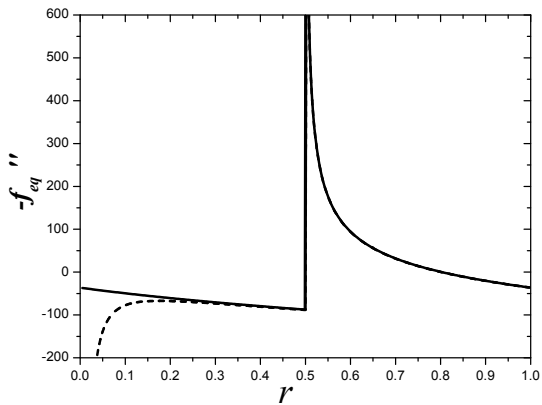}   }

\vspace{2mm}
\caption{
The oscillatory dependence of the magnetic susceptibility $\chi_{eq} \sim -f''_{eq}(r)$
from the equatorial region in the applied magnetic field $H$,
expressed in terms of the dimensionless parameter $r \sim 1/H$ (see text).
The dashed line stands for the contribution from the first equatorial tube, the solid line -- from the
first and second equatorial tubes.
} \label{fig7}
\end{figure}

In calculating the magnetic susceptibility $\chi_{eq}$ one has to keep in mind that
$\triangle E_{eq}$ depends on $H$ through $\omega$ explicitly and on $r$ implicitly.
As shown in Appendix C1, the contribution from the derivative of $r(H)$ with respect to the magnetic field $H$ is dominant.
Finally, we obtain
\begin{eqnarray}
  \chi_{eq} = -\frac{\sqrt{2} m \pi \rho}{15} \;
 \omega^2 \sqrt{\frac{m}{\hbar}\, \omega} \; \frac{\partial^2 f_{eq}(r) }{\partial r^2 } \left( \frac{\partial r}{\partial H} \right)^2 .
 \label{q24}
\end{eqnarray}
The plot of $\chi_{eq}(r)$ is reproduced in Fig.\ \ref{fig7}.
It is worth noting that $\chi_{eq,1}$ diverges at $r \rightarrow 0^+$
(the divergence disappears when the second equatorial tube is accounted for, see below)
and at $r \rightarrow (1/2)^+$.
The latter persists in a more refined calculation with two or more equatorial tubes, because
it is connected with the onset of the occupation of a new Landau level in the $(k_x,k_y)$ equatorial plane.
%
%

Notice that if we limit ourselves to the case of only first equatorial tube, then in correspondence with Eqs.\ (\ref{q22a}) and (\ref{q22b}),
the oscillatory energy contribution at $r=0$ and $r=1$ is different, namely $\triangle E_{eq,1}(r=0) = 0$ $\triangle E_{eq,1}(r=1) \neq 0$,
Fig.\ \ref{fig6}.
In reality the physical situation is the same, the condition $r=0$ simply implies that the first equatorial tube is absent,
while the second equatorial tube plays the role of the first.
The inconsistence exists for the other quantities, for example, for the magnetic susceptibility, Fig.\ \ref{fig7}.
Therefore, to make the values at $r=0$ and $r=1$ consistent, we have to take into account the irregular term
from the second equatorial tube.
Then the contribution from the equatorial region $\triangle E_{eq}$, described by (\ref{q21}), changes,
\begin{eqnarray}
  \triangle E_{eq}=\triangle E_{eq,1}+\triangle E_{eq,2} .
 \label{ne0}
\end{eqnarray}
and the function $f_{eq}(r)$ in (\ref{q21}) becomes
\begin{eqnarray}
  f_{eq}(r)=f_{eq,1}(r)+f_{eq,2}(r) .
 \label{ne0a}
\end{eqnarray}
All necessary calculations are given in Appendix C2 of the Supplementary Materials, and numerical results are shown
by solid lines in Figs.\ \ref{fig6A}, \ref{fig6} and \ref{fig7}.
It is worth noting that except for the range around $r=0$ and $r=1$, the inclusion of the second equatorial tube
plays only a minor role, Figs.\ \ref{fig6A}, \ref{fig6}.
In the magnetic susceptibility plot, Fig.\ \ref{fig7}, though the extended equatorial region leads to the disappearance
of the divergence at $r \rightarrow 0^+$.
The divergence at $r \rightarrow (1/2)^+$ remains because it arises from a Landau level
crossing the Fermi surface in the equatorial region, Fig.\ \ref{fig7}.

\section{Conclusions}
\label{sec:con}

The diamagnetic susceptibility of the free electron gas (Landau diamagnetism) and the oscillatory de Haas - van Alphen contribution
to the magnetic susceptibility from the equatorial region of the Fermi surface are derived analytically at zero temperature
without summation and integration of the free energy terms.
For that the occupied electron states of the Fermi sphere are partitioned in two regions: the first region includes the vast majority of the electron states
inside the Fermi sphere whose energy does not change in an applied magnetic field and
the second region includes a very narrow stepwise region below the Fermi surface whose energy does change in the applied magnetic field.
The partitioning of electron states is imposed by the structure of Landau levels, around which one can introduce
magnetic tubes in the reciprocal space.
Therefore, the Landau diamagnetic response of the free electron gas can be considered as a Fermi surface effect.

While the diamagnetic response is due to the region just below the Fermi surface, the oscillatory behaviour of
energy and magnetic susceptibility arises from its equatorial part.
We also show that a small oscillatory change of the Fermi energy in the applied magnetic field
is caused by redistribution (inflow or outflow) of electrons from the equatorial region
of the Fermi surface.

Based on the ground state structure considered in this paper it is possible to extend it to the case of finite temperatures
by considering thermal excitations of one dimensional electron states lying on the Landau levels close to the Fermi energy.


\end{document}